# Human peripheral blur is optimal for object recognition

Pramod R T[1]*[#], Harish Katti[2]*[#], & Arun S P[2][#]

[1]Department of Electrical Communication Engineering & [2]Centre for Neuroscience

Indian Institute of Science, Bangalore 560012

*Both authors contributed equally

[#]Correspondence to:

pramodrt9@gmail.com, harish2006@gmail.com & sparun@iisc.ac.in

# ABSTRACT

Our vision is sharpest at the center of our gaze and becomes progressively blurry into the periphery. It is widely believed that this high foveal resolution evolved at the expense of peripheral acuity. But what if this sampling scheme is actually optimal for object recognition? To test this hypothesis, we trained deep neural networks on "foveated" images with high resolution near objects and increasingly sparse sampling into the periphery. Neural networks trained using a blur profile matching the human eye yielded the best performance compared to shallower and steeper blur profiles. Even in humans, categorization accuracy deteriorated only for steeper blur profiles. Thus, our blurry peripheral vision may have evolved to optimize object recognition rather than merely due to wiring constraints.

# INTRODUCTION

Our retina contains 100 times more photoreceptors at the center compared to the periphery (Curcio and Allen, 1990; Curcio et al., 1990). It is widely believed that this sampling scheme saves on the metabolic cost of processing orders of magnitude more information that would result from full resolution scenes without affecting overall performance (Weber and Triesch, 2009; Akbas and Eckstein, 2017). But what if this sampling scheme evolved to optimize object recognition?

There are several lines of evidence suggesting that peripheral blurring can benefit object recognition. First, object detection by humans on natural scenes is slowed down by clutter as well as by partial target matches in the background (Katti, Peelen, & Arun, 2017). Thus, high spatial frequency information in the periphery interferes with recognition. Second, the surrounding scene context can facilitate recognition (Li et al., 2002; Bar, 2004; Davenport and Potter, 2004) but this information is contained in low spatial frequency (Morrison and Schyns, 2001; Torralba, 2003; Bar, 2004; Torralba et al., 2006). Thus, low spatial frequencies in the surround can facilitate recognition. Thus, sampling images densely near objects and sparsely in the surrounding context might be optimal for recognition.

To further illustrate how such a sampling scheme benefits recognition, consider the example scene in Figure 1A. When this scene is given as input to a state-of-the-art pre-trained deep neural network (R-CNN; see Methods), it correctly identified the person but made a false alarm to a traffic cone in the background. We then "foveated" the image by resampling it at full resolution on the salient object (the person) and sampling it sparsely into the periphery according to the human blur function. The same deep network no longer showed the false alarm (Figure 1B). Thus, peripheral blurring can be beneficial in avoiding spurious target matches far away from objects of interest.

Note that foveating on the foreground object or objects does not by itself yield enough information about object identity.

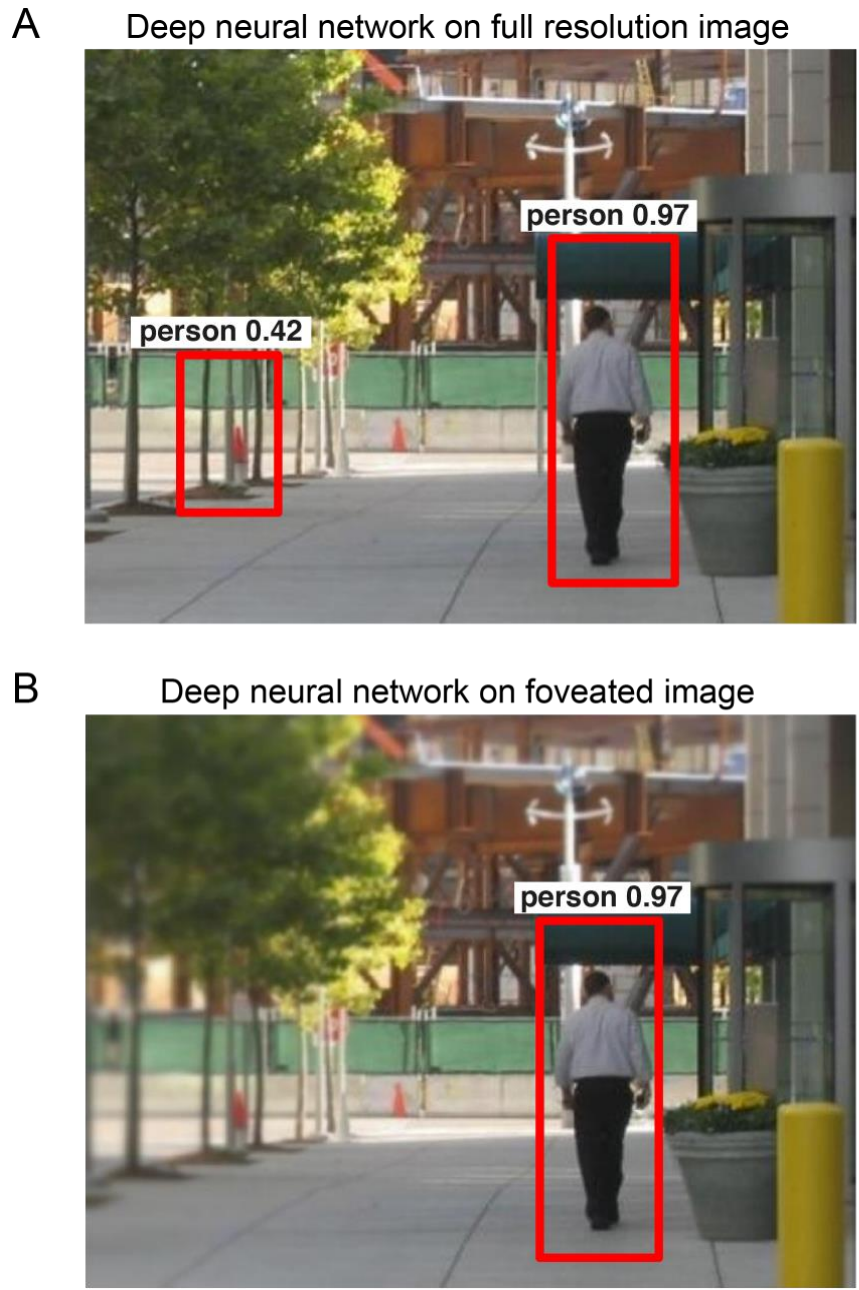

**Figure 1. Example object detection with and without peripheral blur.**
- **(A)** Example object detections from a state-of-the-art deep neural network (R-CNN), showing a correctly identified person and a false alarm in which a traffic cone is mistaken for a person.
- **(B)** Example object detections on a foveated version of the image using the same network, showing the correctly identified person but without the false alarm.

# RESULTS

If peripheral blur is optimal for recognition, then it follows that object classifiers trained on foveated images (with high resolution near objects and gradual blur into the periphery) should progressively improve recognition until performance peaks for the human peripheral blur profile. We tested this hypothesis by training state-of-the-art deep neural network architectures on foveated images with varying peripheral blur profiles.

To train these deep networks, we selected images from the widely used ImageNet dataset (~500,000 images annotated with object category and location across 1,000 object categories). Importantly, these images are photographs taken by humans in a variety of natural viewing conditions, making them roughly representative of our own visual experience. To obtain foveated images, we started with the well-known human contrast sensitivity function measured at different eccentricities from the fovea (Geisler and Perry, 1998). To apply this peripheral blur correctly in degrees of visual angle, we assumed that images are viewed at a distance of 120 cm. We obtained qualitatively similar results on varying this choice (Section S1).

To vary the peripheral blur profile, we fitted this function to an exponential and modified its spatial decay by a factor of 0.5, 1, 2 or 4 (Figure 2A). We then applied this blur profile to each image, centred on the labelled object (see Methods). Example images with varying degrees of peripheral blur are shown in Figure 2. A spatial decay factor smaller than 1 indicates shallower decay than human peripheral blur, i.e. the image is in high resolution even into the periphery (Figure 2B). A value of 1 indicates images blurred according to the human peripheral blur function (Figure 2C). A value larger than 1 indicates steeper decay i.e. the image blurs out into the periphery much faster than in the human eye (Figure 2D).

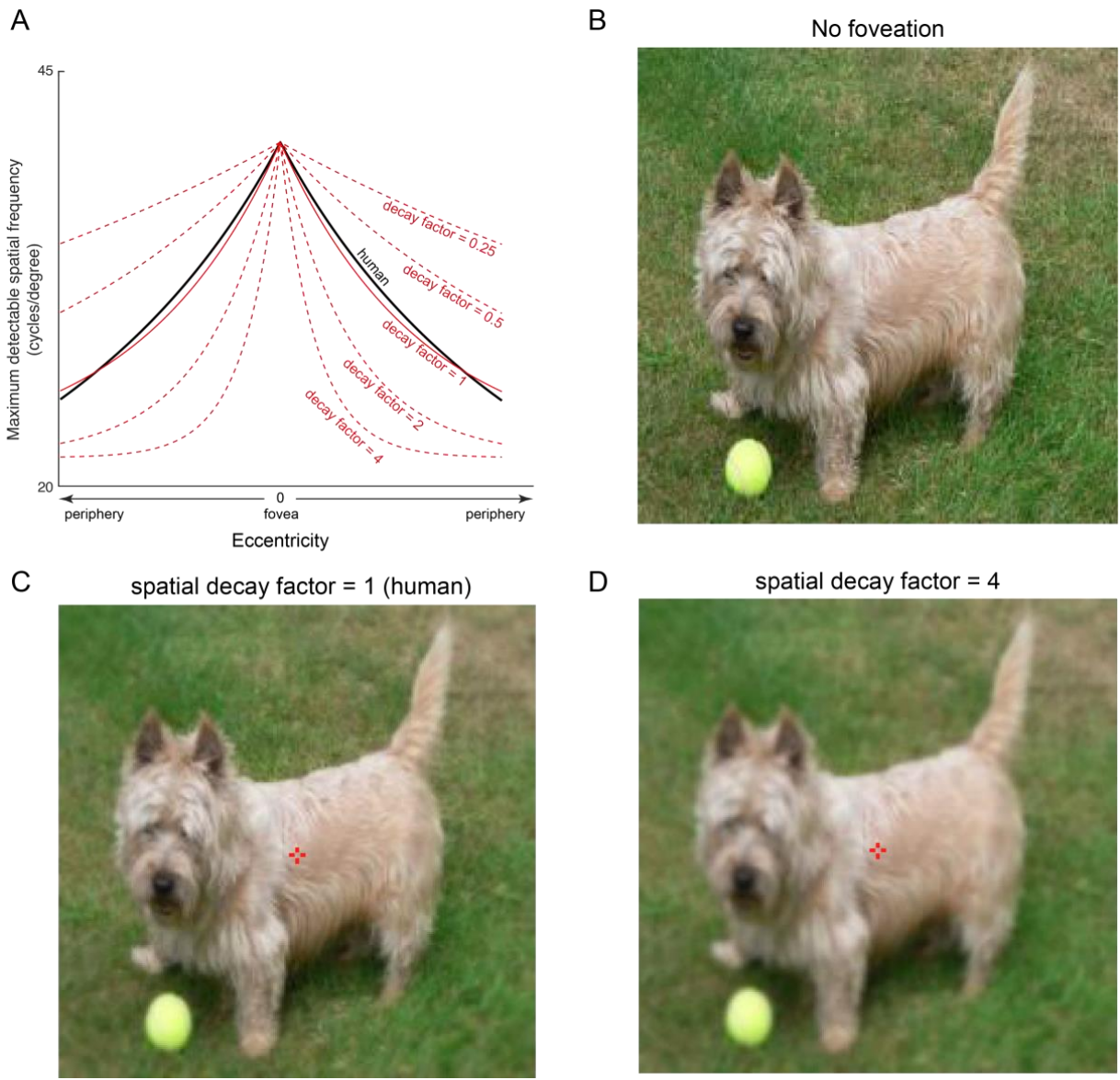

**Figure 2. Example foveated images with varying peripheral blur profiles.**
 (A) Human contrast sensitivity function (*solid black line*) and the corresponding exponential fit (*solid red line*). The spatial decay of the exponential was varied by scaling the human exponential fit to obtain shallower or deeper blur profiles (*dashed red lines*).
 (B) Example full resolution image
 (C) Same as panel B but foveated on the object center (*red cross*) with a spatial decay of 1, which corresponds to the human peripheral blur function. At the appropriate viewing distance, fixating on the red cross will make this image look identical to the full resolution image in panel B.
 (D) Same as panel B but foveated with a more extreme peripheral blur (spatial decay factor = 4).

**Foveation leads to increased object recognition performance**

Next we trained a widely used deep convolutional neural network architecture (VGG-16; Figure 3A) for 1000-way object classification on the ImageNet images with bounding box annotations. We trained a total of six neural networks: one network was trained on full resolution images (no foveation) while the other five networks were trained on images with different degrees of foveation with spatial decay factors of 0.25, 0.5, 1, 2 and 4 (Figure 3B). We then tested each network for its generalization abilities by evaluating its classification accuracy on novel images foveated with the corresponding spatial decay factor. Their performance is summarized in Table 1.

| spatial decay factor | Top-1 accuracy | | Top-5 accuracy | |
| --- | --- | --- | --- | --- |
| | Train | Test | Train | Test |
| 4 | 58.9 | 48.0 | 81.0 | 72.0 |
| 2 | 63.4 | 49.7 | 83.0 | 73.4 |
| **1 (human)** | **73.0** | **52.1** | **89.0** | **75.5** |
| 0.5 | 64.8 | 50.7 | 84.8 | 74.0 |
| 0.25 | 61.0 | 49.0 | 82.0 | 72.0 |
| 0 (No foveation) | 65.9 | 50.0 | 84.6 | 73.9 |

**Table 1. Classification performance of VGG-16 networks on foveated and full resolution images.** We report both Top-1 and Top-5 accuracies on both training and test sets. The Top-1 accuracy refers to the accuracy with which the best guess of the network matched the object label. The Top-5 accuracy is the accuracy with which the correct label was present in the top 5 guesses of the network. The network trained on images with human-like peripheral blur (spatial decay factor = 1) is highlighted in *red*.

Across all networks, the network trained on images foveated according to the human peripheral blur function gave the best performance (Top-1 accuracy = 52.1% and Top-5 accuracy = 75.5%; Figure 3B; Table 1). This performance was significantly better than the network trained on full-resolution images (Increase in top-1 accuracy: mean ± std: 2% ± 0.9% across 1000 categories; $p < 0.000005$, signed-rank test; increase in top-5 accuracy, mean ± std: 1.66% ± 0.25%, $p < 0.000005$, signed-rank test). Thus, using human-like peripheral blur yielded optimal recognition performance compared to other blur profiles.

To investigate the reason behind the improved performance of the network trained on foveated images, we reviewed images that were correctly classified after foveation but were misclassified without foveation (Figure 3C). We observed two types of benefits. First, foveation helped to disambiguate between similar categories, such as in the "digital watch" and "freight car" images. Here, the full-resolution network incorrectly classified these images as "digital clock" and "passenger car" but the foveated network correctly classified them. Likewise the "airliner" is classified as "war plane" and "spacecraft" with higher probability than "airliner" itself by the full-resolution network but is correctly classified after foveation. Second, foveation improved the quality of top-ranked guesses as in the case of "dalmatian" where the full-resolution network determined other categories as more likely (trilobite, hook, necklace). The foveated network also made reasonable guesses for the other likely categories (Great Dane, English Foxhound, etc).

Next we wondered whether these results would generalize to other image datasets or neural network architectures. To this end we trained another neural network architecture for person categorization over images chosen from the MSCOCO database (Lin et al., 2014). Here too, using human-like peripheral blur yielded optimal performance (Section S2).

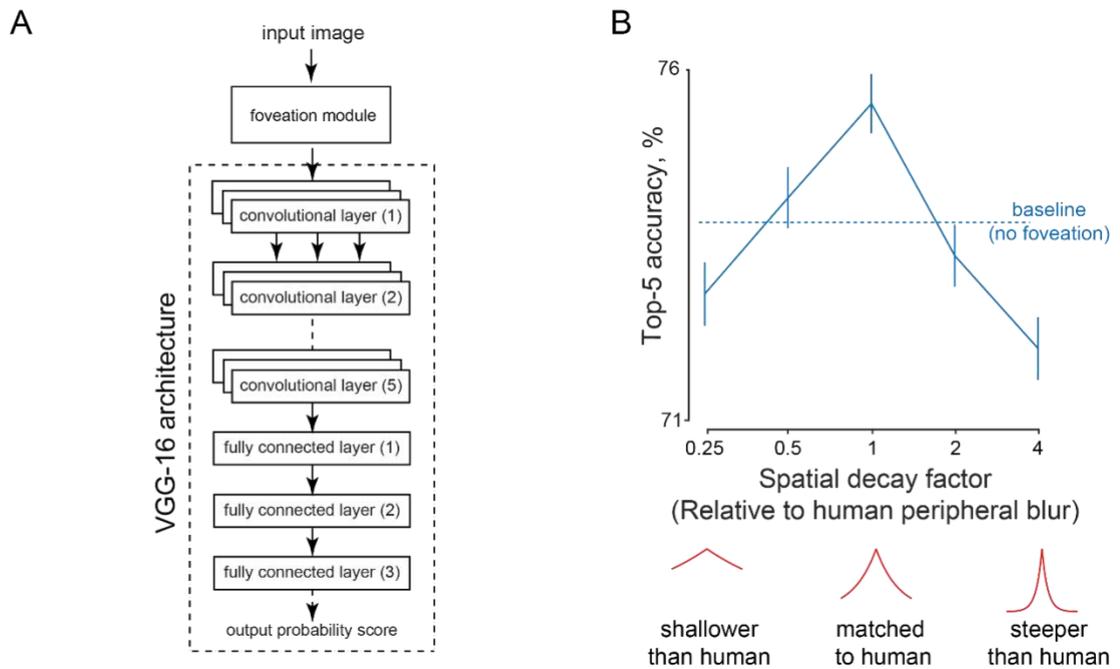
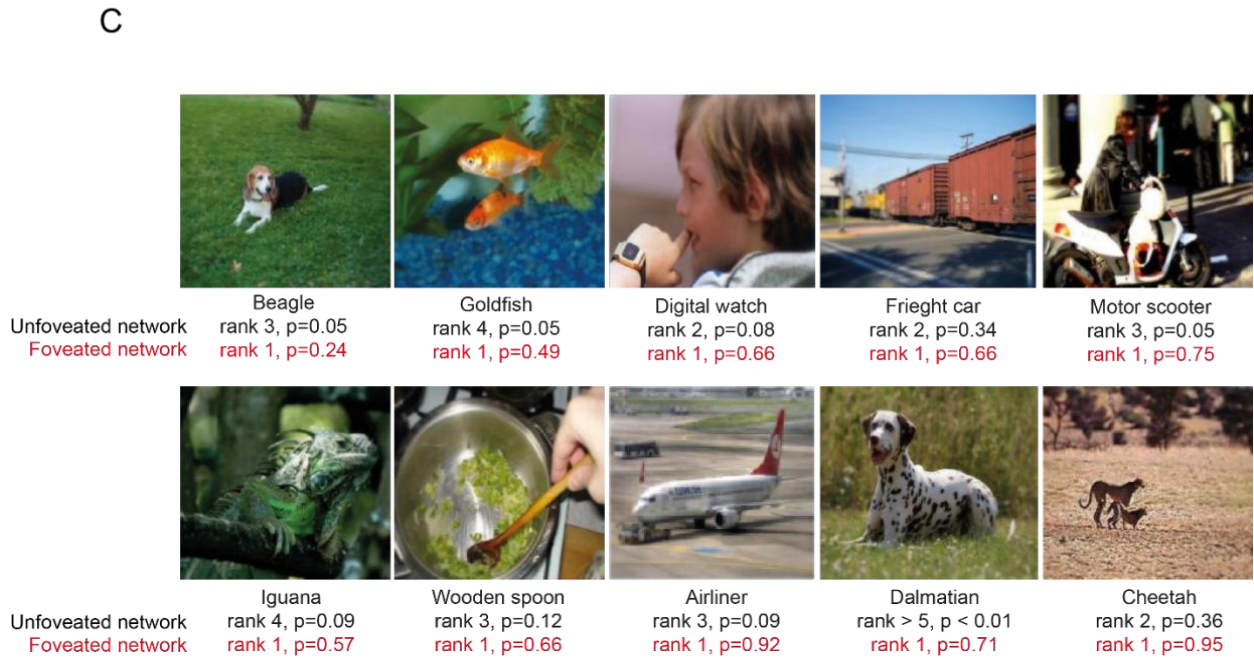

**Figure 3. Human-like peripheral blur is optimal for object recognition.**
  (A) Schematic of the VGG-16 neural network architecture used to train images
  (B) Top-5 accuracy of neural networks with varying peripheral blur. The accuracy of each network was calculated on test images after training it on foveated images with the corresponding blur profile. Baseline accuracy for the network trained on full-resolution images is shown using dotted lines.
  (C) Example images for which the correct category was identified only by the foveated network (with human-like peripheral blur) but not the full-resolution (unfoveated) network. Below each image, the correct object label is shown (*top*), followed by its rank and posterior probability returned by the unfoveated network (*black, second row*) and by the foveated network (*red, third row*).

**Evolution of the foveation advantage across neural network training**

In the above results, the overall improvement of the network with human-like foveation could arise from improved detection of objects, or a decrease in the rate of false alarms. It could also arise early or late during training which may further elucidate the nature of the underlying features. To investigate this possibility, we saved the model weights every five epochs during training and calculated the overall percentages of hits and false alarms. We then calculated hits and false alarms over the course of learning for two networks: the best network (with human-like foveation) and the network trained on full resolution images (*no foveation*). We found that the improvement in accuracy for the foveated network largely came from both an increase in the hits (Figure 4A) and a reduction in false alarms (Figure 4B). This trend emerged very early during network training and remained consistent through the course of training. Thus, the network trained on foveated images achieves greater accuracy fairly early on during training and learns faster.

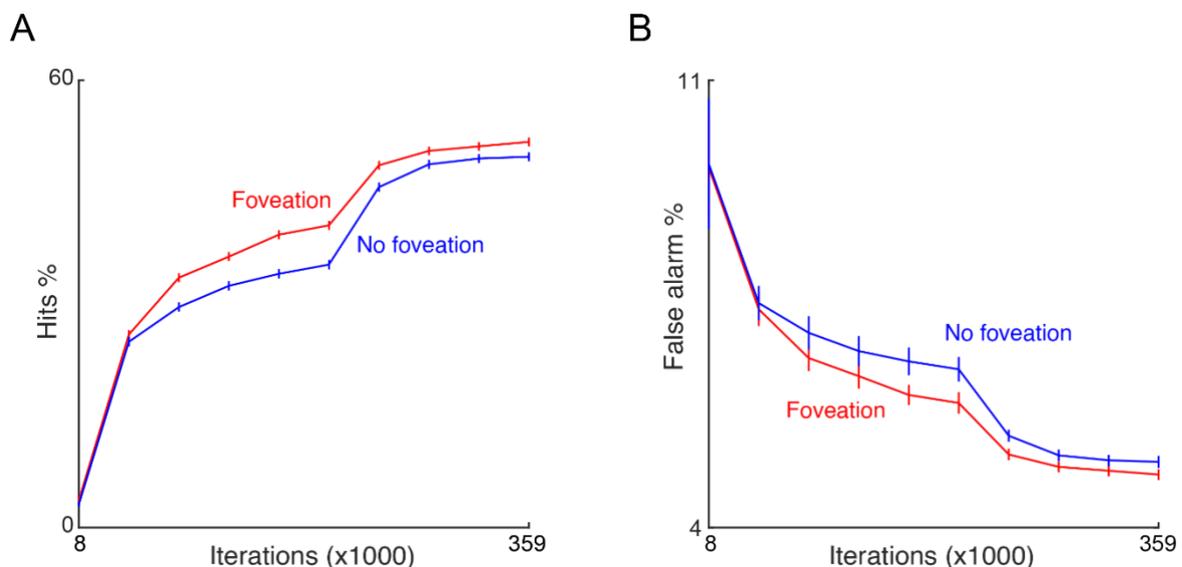

**Figure 4. Object recognition performance over the course of training.** (A) Plot of percentage hits as a function of learning for networks trained on foveated images (*red*) and full resolution images (*blue*). (B) Same as in (A) but for false alarms. In both plots the x-axis indicates the number of iterations (or batches of data) in multiples of 1000. Error bars indicate s.e.m. across 1000 categories.

**Evaluation of relevant spatial information**

The above results demonstrate that human-like foveation is optimal for object recognition. This raises the intriguing possibility that foveation in the eye may have evolved to optimize object classification. Did this evolution require a complex neural network architecture, or could it arise from simpler feature detectors? To examine this possibility, we wondered whether the image features most useful for recognition vary progressively with distance from the object in a scene. Specifically, we predicted that the low spatial frequency information is more discriminative for object recognition at peripheral locations whereas high spatial frequency information is more relevant at the fovea. If this is true, then even simple classifiers based on spatial frequency features could potentially drive the evolution of foveal vision.

To verify this, we selected a subset of 11 categories from the ImageNet validation dataset. For each image, we extracted image patches at varying distances from the center and used a bank of Gabor filters to extract low and high spatial frequency filter responses from each image patch. We then trained linear classifiers on the responses of each spatial frequency filter to image patches at a particular distance from the centre. The results are summarized in Figure 5A.

Object decoding accuracy was significantly higher than the chance performance (1/11 = 9%) at all eccentricities and all spatial frequencies, indicating that there is object-relevant information at all locations and frequencies (Figure 5). However, it can be seen that classification accuracy was best for high spatial frequency features at the center, and best for low spatial frequency into the periphery. Thus, even simple detectors based on spatial frequency features show an advantage for sampling densely at the center and sparsely in the periphery.

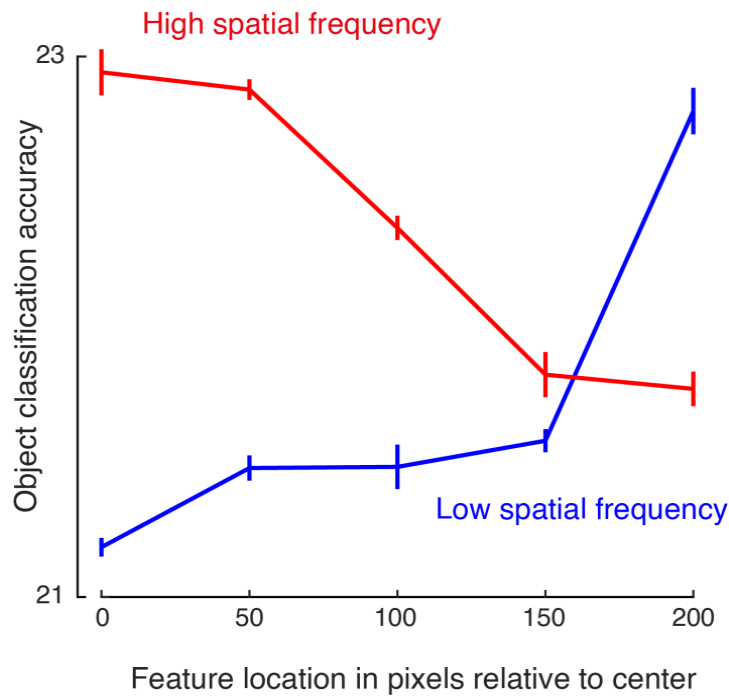

**Figure 5. Relative importance of spatial frequency features as a function of image eccentricity.** Accuracy of a 11-way object decoder is plotted as a function of eccentricity i.e. feature location in pixels relative to the image center, for high spatial frequencies (*red*) and low spatial frequencies (*blue*).

**Human categorization on foveated images**

Our finding that human-like foveation is optimal for recognition is based on training neural networks. We therefore wondered how image categorization by humans would change across varying peripheral blur profiles. Since human eyes are already equipped with the typical peripheral blur profile, we predicted that foveating images with spatial decay factor of less than 1 should have no effect on recognition performance. Further, viewing images with steeper blur profiles should lead to reduced performance, due to the lack of useful low-frequency features in the periphery.

We evaluated these predictions using several behavioural experiments on humans. In Experiment 1, subjects had to indicate whether briefly presented scene contained an animal or not (see Methods). An example image is shown in Figure 6A. We used four types of images: full resolution and three levels of foveation with spatial decay factors of 0.25, 1 and 4. Critically, to avoid memory effects, subjects saw a given scene only once across all levels of foveation.

Subjects were highly accurate on this task (accuracy, mean ± std: 94% ± 1.1% across the four types of images). Importantly, accuracy was significantly lower for steeply foveated images (spatial decay factor = 4) compared to other variants (average accuracy: 93% for steeply foveated images and 94.9%, 94.6% and 94.5% for full resolution, and images with spatial decay factors of 0.25 and 1 respectively; $p < 0.005$ for ranksum test on average accuracies for foveated images with spatial decay factor of 1 vs 4; Figure 6B). Further, subjects' accuracy was comparable for full resolution images and human foveated images with spatial decay factor of 1 ($p = 0.29$ using ranksum test on average accuracies across images).

We found similar but stronger effects of foveation on reaction times. Reaction times were slowed down only for the highest spatial decay factor (reaction times, mean ± std: 529 ± 102 ms, 523 ± 93 ms, 527 ± 95 ms and 545 ± 98 ms for full resolution images, and foveated images with spatial decay factors of 0.5, 1 and 4 respectively; $p < 0.0005$ for ranksum test on reaction times for human foveated and steep foveated images, $p > 0.05$ for all other pairwise comparisons; Figure 6C).

Next, we asked whether these results would generalize to other categories. To this end, in Experiment 2, subjects had to detect the presence of people in an image (example scene in Figure 6D). Subjects were highly accurate in detecting the target object (accuracy, mean ± std across subjects: 75.3% ±1.5% across the four types of images). As with the animal categorization task, accuracy was lowest for steeply foveated images (average accuracy: 76% for full resolution; 76.8% for shallow foveation; 75.4% for human foveation; 73.3% for steep foveation; Figure 6E).

We found similar but stronger effects in reaction times. Reaction times were the slowest for steeply foveated images (reaction times, mean ± std: 566 ± 182 ms, 547 ± 199 ms, 558 ± 374 ms and 577 ± 215 ms for full resolution images, and foveated images with spatial decay factors of 0.5, 1 and 4 respectively; p = 0.009 for ranksum test on reaction times for human foveated and steep foveated images; p = 0.02 for ranksum test on reaction times for full-resolution and human foveation; p > 0.05 for all other pairwise comparisons; Figure 6F).

Finally, to verify whether this effect is specific to animate objects, we performed an additional experiment in which subjects performed car detection. Here too, we observed similar results (Section S3).

To summarize, categorization performance in humans remained similar for both full resolution and foveated images, and worsened only for steeper levels of foveation. This is expected because humans already have peripheral blur in their eyes, as a result of which only steep foveation has any impact on performance.

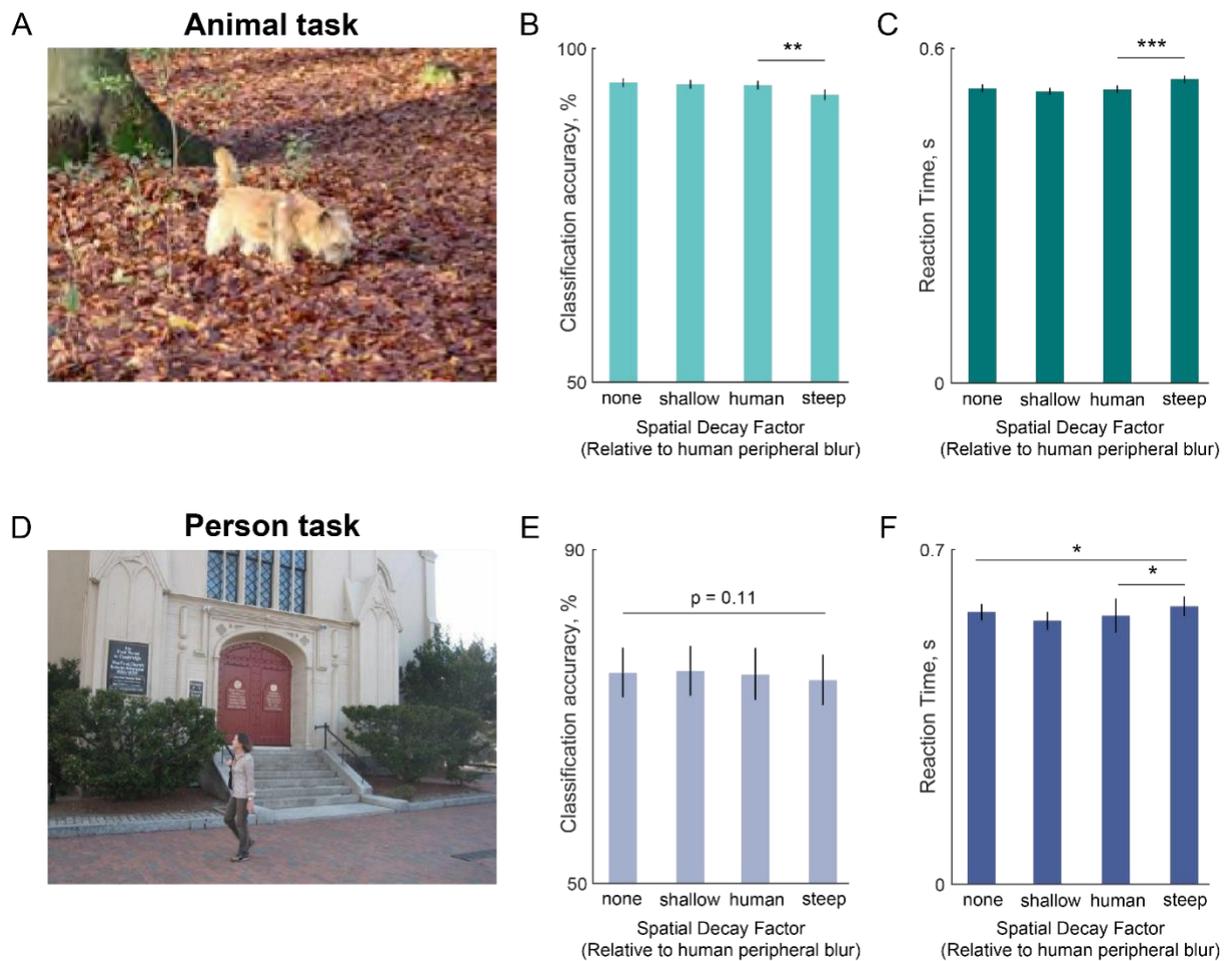

**Figure 6. Human categorization declines only for steep foveation.**
  (A) Example full resolution image from the animal categorization task.
  (B) Accuracy for different levels of foveation. Error bars indicate s.e.m. calculated across all images used in the task. Asterisks indicate statistical significance using a Wilcoxon signed rank-sum test across images (* is p < 0.05, ** is p < 0.005, *** is p < 0.0005). All other comparisons are not significant.
  (C) Same as (B) but for reaction times on correct trials, with error bars indicating s.e.m across images. Conventions are as in (B).
  (D) Example full resolution image from the person categorization task.
  (E-F) Same as (B) and (C), but for the person categorization task.

# DISCUSSION

Our vision is sharpest at the center of gaze and blurs out into the periphery. The coarse sampling of the periphery is widely thought to save on wiring and metabolic cost without impacting performance. Our results challenge this belief by showing that the human peripheral blur profile is actually optimal for object recognition on natural images. This in turn implies that the evolution of a fovea might have been driven by the demands of visual recognition rather than to simply satisfy wiring constraints.

Our specific findings in support of this conclusion are: (1) Deep networks trained on natural images show optimal performance for human-like foveation; (2) The relevant features for object recognition require high spatial frequencies near the image center and low spatial frequencies in the periphery; and (3) Humans performing categorization on natural scenes show a decline in categorization only when scenes are blurred beyond the normal level of peripheral blur. Below we discuss these findings in the context of the relevant literature.

Our main finding is that deep networks trained on foveated images achieve optimal performance for human-like peripheral blur (Figure 3). This raises several important concerns that merit careful consideration. First, could this improvement come from the foreground object becoming more salient with peripheral blurring? We consider it unlikely because this would predict a monotonic increase in accuracy with steeper blur profiles, which is opposite to what we observed. Second, if full-resolution images contain more information than foveated images, then why do deep networks achieve lower accuracy on full-resolution images? This could be because full-resolution images contain target-like features in the periphery that result in false alarms or slow detection (Katti et al., 2017). It could also be that deep networks trained on full-resolution images fail to pick up important scene context features (Zhu et al., 2016; Katti et al., 2019). Third, if foveation is optimal for recognition, then how does the visual system know where to foveate before initiating recognition? There is a large body of evidence showing that the primate oculomotor system uses a saliency map to guide saccades, and that low-level features can be used to guide eye movements towards potential objects of interest (Itti and Koch, 2001; Akbas and Eckstein, 2017). Whether and how the ventral stream visual regions influence the saliency map can be elucidated through paired recordings in both regions.

The finding that human-like peripheral blur yields optimal recognition in deep networks alone does not constitute proof that human peripheral blur evolved to optimize recognition. However, it is a remarkable coincidence that the exact human peripheral blur profile is what ends up being optimal for recognition. It could be argued that feature detectors in our brains are qualitatively different from deep networks, but there is growing evidence that this is not the case: object representations in deep networks have strong parallels to the ventral visual stream neural representations (Yamins et al., 2014; Ponce et al., 2019).

Our conclusion that foveation might have evolved for optimal recognition stands in stark contrast to the literature. Previous studies have used foveation as a pre-processing step to achieve image compression (Geisler and Perry, 1998) or to create saliency maps to guide eye movements (Itti and Koch, 2001). However no previous study has systematically varied peripheral blur profiles to examine the impact on recognition. A recent study has shown that foveation yields equivalent object detection performance to full-resolution images but with significant computational cost savings (Akbas and Eckstein, 2017). If foveation is so beneficial for object recognition, then why has this not been noticed previously? In our experiments, we observed consistently better performance for foveated images, but this benefit varied with the viewing distance used in the foveation calculations. We speculate that these studies may have used sub-optimal values of viewing distance, resulting in only marginal improvements.

We have shown that low-spatial frequency features are most informative for object detection in the image periphery, whereas high-spatial frequency features are most informative at the image center. These results are concordant with the recent observation that a fovea-like sampling lattice evolves after training a deep network for handwritten digit recognition (Cheung et al., 2017). These findings suggest that the evolution of a fovea can be driven by object detectors based on simple Gabor-like features as have been observed in the primary visual cortex. More generally, we note that the organization of the fovea varies widely across animals (Land and Nilsson, 2012). We speculate that the fovea and peripheral blur profile in each species may be optimized for its high-level visual demands, just as our eyes are optimized to ours.

## METHODS

**Generating foveated images**

Any visual stimulus can be analysed in terms of its spatial frequency content with fine details (like edges) attributed to high spatial frequencies and coarse information (like object shape) attributed to low spatial frequencies. The range of visible spatial frequencies is usually measured as the sensitivity to contrast at each spatial frequency and is summarized by the contrast sensitivity function (CSF) which varies as a function of retinal eccentricity (Campbell and Robson, 1968). Based on previous research using grating stimuli for simple detection/discrimination tasks, the contrast threshold for detecting a grating patch of spatial frequency *f* at an eccentricity *e* is given by

$$CT(f,e) = CT_0 \exp\left(\alpha f \frac{e+e_2}{e_2}\right) \quad (1)$$

where *f* is spatial frequency (cycles per degree), *e* is the retinal eccentricity (degrees), $CT_0$ is the minimum contrast threshold, α is the spatial frequency decay constant, and $e_2$ is the half-resolution eccentricity. We took the values of these variables to be $CT_0$ = 0.0133, α = 0.106, $e_2$ = 2.3 respectively. This formula matches contrast sensitivity data measured in humans under naturalistic viewing conditions (Geisler and Perry, 1998). Although the above formula gives the contrast threshold, what is more important is the critical eccentricity $e_c$ beyond which the spatial frequency *f* will be invisible no matter the contrast. This critical eccentricity for each such spatial frequency *f*, can be calculated by setting the left-hand side of the equation above to 1 and solving for *e*.

$$e_c = \frac{e_2}{\alpha f} \ln\left(\frac{1}{CT_0}\right) - e_2 \quad (2)$$

The above equation for critical eccentricity (in degrees) was then converted to pixel units by considering the viewing distance. Specifically, critical eccentricity in cm is calculated using the formula

$$e_{c,cm} = d * \tan\frac{\pi e_c}{180} \qquad (3)$$

where $e_{c,cm}$ is the critical eccentricity beyond which spatial frequency *f* (equation 2) will be unresolvable at a viewing distance *d* (in cm) (see below for choice of *d*). This was then converted into pixel units using dot-pitch of the monitor (in cm).

$$e_{c,px} = \frac{e_{c,cm}}{pitch} \qquad (4)$$

The dot-pitch value of the monitor in our experiments was 0.233 cm. Then, the input image was low-pass filtered and down-sampled successively by a factor of two, to create a multi-resolution scale pyramid having up-to seven levels. Further, *f* in the above equation for $e_c$ was set to be the Nyquist frequency at each level of the multi-resolution scale pyramid and the resulting values of $e_c$ were used to define the foveation regions at each level. That is, pixel values for the foveated image were chosen from different levels of the multi-resolution scale pyramid according to the eccentricity of the pixel from the point of fixation. In our experiments, in addition to using the default values of all the parameters, we obtained different foveation blur profiles by modulating α by a spatial decay factor γ.

$$\alpha_{new} = \alpha\gamma \text{ for } \gamma = \{0.25, 0.5, 1, 2, 4\} \qquad (5)$$

where γ is the spatial decay factor with γ = 1 being the human foveation blur profile (Equation 1).

**Example object detection with and without peripheral blur**

To illustrate object detection with and without peripheral blur, we took a pre-trained deep neural network (Faster R-CNN) that yields state-of-the-art performance on object detection (Ren et al., 2015). This network had been pre-trained to identify instances of 20 different classes including people. To this neural network we gave as input both the full-resolution scene as well as the foveated image with human peripheral blur. The resulting object detections for the "person" class are depicted in Figure 1.

**CNN training: VGG-16 architecture trained on ImageNet with foveation**

To test if foveation is computationally optimal for object recognition in natural scenes, we chose ~500,000 images from the ImageNet dataset with manual object level bounding box annotations. We created 5 foveated versions of each image with the point of foveation fixed at the centre of the bounding box and trained deep neural networks for object recognition. Specifically, we used VGG-16 architecture and trained six separate networks (one for the full resolution and five for different foveated versions of the image). Note that, all foveated images were created after scaling the image to 224x224 pixels which is the default size of input to the VGG-16 network. To create images with different levels of foveal blur, we used the equations described in the previous section. The output of those equations depends crucially on the distance between the observer and the image.

How do we find the viewing distance for the deep network? To estimate the optimal viewing distance, we trained separate networks on images foveated with a viewing distance of 30, 60, 90, 120 and 150 cm. We obtained consistent improvements in performance for all choices of viewing distance, but the best performance was obtained for a viewing distance of 120 cm. We used this value for all the reported analyses. However we confirmed that our results are qualitatively for other choices of viewing distance (Section S1).

For each network, we started with randomly initialized weights and trained the network for 1000-way object classification over 50 epochs of the data with a batch-size of 32. All networks were defined and trained using the PyTorch framework with NVIDIA TITAN-X/1080i GPUs. All the trained models were tested for generalization capabilities on a corresponding test set containing 50,000 images (ImageNet validation set).

**Evaluation of spatial frequency content**

To explore the relationship between spatial frequency content and object recognition, we selected 11 random categories from the ImageNet validation dataset - these were categories 1:100:1000 from ImageNet, which included common objects like fish, bird, animal, insect, clothing, building etc. We rescaled all images to have at least 500 pixels along both dimensions and chose 100 pixels x 100 pixels patches on concentric circles with radii 0, 50, 100, 150 and 200 pixels from the centre of the image. These patches were chosen along 8 equally spaced directions on the circle with the exception of the patch at the centre which was considered only once. We then extracted low and high spatial frequency from a bank of Gabor filters tuned for six spatial frequencies (0.06, 0.09, 0.17, 0.25, 0.33 and 0.5 cycles/pixel) and 8 orientations (uniformly sampled between 0 and 180 degrees). We then trained linear object identity decoders at both foveal as well as peripheral locations on the concatenated filter responses across all patches corresponding to high or low spatial frequencies.

**Experiment 1: Animal detection task**

All experiments were conducted in accordance to an experimental protocol approved by the Institutional Human Ethics Committee of the Indian Institute of Science. Subjects had normal or corrected-to-normal vision, gave written informed consent and were monetarily compensated for their participation.

*Subjects.* A total of 58 subjects (18-52 years, 22 females) participated in this experiment.

*Procedure.* Subjects were comfortably seated ~60 cm from a computer monitor with a keyboard to make responses. Image presentation and response collection was controlled by custom scripts written in MATLAB using Psychtoolbox (Brainard, 1997). Each trial began with a fixation cross at the centre of the screen shown for 500ms followed by the image. All images measured 640 x 480 pixels, and subtended 13.5° in visual angle along the longer dimension. Images were shown at the centre of the screen for 100 ms followed by a white-noise mask. The noise mask stayed for 5 s or till the subject responded, whichever was earlier. Subjects were instructed to respond as quickly and as accurately as possible to indicate whether the image contained an animal or not ('a' for animals and 'n' otherwise).

*Stimuli.* We created three groups of foveated images with spatial decay factors of 0.25, 1 and 4. For each group, we chose 212 full resolution images of animals (for example: duck, dog, elephant, fowl, deer, rabbit, ostrich, buffalo) and an equal number of images of inanimate objects (for example: boat, bicycle, wheelbarrow, airplane, flowerpot, tower, hot air balloon, letterbox, car). All images were chosen from the ImageNet validation set. The retinal sizes of key objects in the animate and inanimate categories were comparable. In all, there were 1696 images (424 images of animals and inanimate objects x 4 levels of foveation). Subjects saw 424 images (212 each of animals and inanimate objects) such that each image was shown in only one of the foveated conditions. This was achieved by dividing the set of 212 category images into 4 mutually exclusive subsets each with 53 images and picking one of these subsets for presentation. We repeated this procedure for all versions (one full resolution and three foveated) and chose non-overlapping subsets of images across versions for the experiment. Each subject saw 424 images, and a given image was shown to 14 subjects.

**Experiment 2: Person detection task**

All methods were identical to Experiment 1, except for those detailed below.

*Subjects:* A total of 31 subjects (18-36 years, 12 female) participated in the task.

*Stimuli:* We chose 120 images that contained people and 120 images that had other objects (for example: dog, bird, boat, dustbin, post-box, bench, window, chair). These images were chosen from the publicly available MS-COCO (Lin et al., 2014) and ImageNet (Russakovsky et al., 2015) datasets. Like in the animal task, we generated three foveated versions of each image wherein one version had a foveal blur that matched the human contrast sensitivity function and two additional ones that were shallower or steeper (spatial decay factors 1, 0.25 or 4). Every participant saw a given scene only once across all versions.

# REFERENCES

Akbas E, Eckstein MP (2017) Object detection through search with a foveated visual system Einhäuser W, ed. PLOS Comput Biol 13:e1005743.

Bar M (2004) Visual objects in context. Nat Rev Neurosci 5:617–629.

Brainard DH (1997) The Psychophysics Toolbox. Spat Vis 10:433–436.

Campbell FW, Robson JG (1968) Application of fourier analysis to the visibility of gratings. J Physiol 197:551–566.

Cheung B, Weiss E, Olshausen B (2017) Emergence of foveal image sampling from learning to attend in visual scenes. In: 5th International Conference on Learning Representations, ICLR 2017 - Conference Track Proceedings.

Curcio CA, Allen KA (1990) Topography of ganglion cells in human retina. J Comp Neurol 300:5–25.

Curcio CA, Sloan KR, Kalina RE, Hendrickson AE (1990) Human photoreceptor topography. J Comp Neurol 292:497–523.

Davenport JL, Potter MC (2004) Scene consistency in object and background perception. Psychol Sci 15:559–564.

Geisler WS, Perry JS (1998) Real-time foveated multiresolution system for low-bandwidth video communication. Proc SPIE 3299:294–305.

Itti L, Koch C (2001) Computational modelling of visual attention. Nat Rev Neurosci 2:194–203.

Katti H, Peelen M V, Arun SP (2017) How do targets, nontargets, and scene context influence real-world object detection? Atten Percept Psychophys 79:2021–2036.

Katti H, Peelen M V, Arun SP (2019) Machine vision benefits from human contextual expectations. Sci Rep 9:2112.

Land MF, Nilsson D-E (2012) Animal eyes, 2nd Editio. New York, NY: Oxford University Press.

Li FF, VanRullen R, Koch C, Perona P (2002) Rapid natural scene categorization in the near absence of attention. Proc Natl Acad Sci U S A 99:9596–9601.

Lin T-Y, Maire M, Belongie S, Bourdev L, Girshick R, Hays J, Perona P, Ramanan D, Zitnick CL, Dollár P (2014) Microsoft COCO: Common Objects in Context. arXiv.

Morrison DJ, Schyns PG (2001) Usage of spatial scales for the categorization of faces, objects, and scenes. Psychon Bull Rev 8:454–469.

Ponce CR, Xiao W, Schade PF, Hartmann TS, Kreiman G, Livingstone MS (2019) Evolving Images for Visual Neurons Using a Deep Generative Network Reveals Coding Principles and Neuronal Preferences. Cell 177:999-1009.e10.

Russakovsky O, Deng J, Su H, Krause J, Satheesh S, Ma S, Huang Z, Karpathy A, Khosla A, Bernstein M, Berg AC, Fei-Fei L (2015) ImageNet Large Scale Visual Recognition Challenge. Int J Comput Vis 115:211–252.

Torralba A (2003) Contextual priming for object detection. Int J Comput Vis 53:169–191.

Torralba A, Oliva A, Castelhano MS, Henderson JM (2006) Contextual guidance of eye movements and attention in real-world scenes: the role of global features in object search. Psychol Rev 113:766–786.

Weber C, Triesch J (2009) Implementations and Implications of Foveated Vision. Recent Patents Comput Sci 2:75–85.

Yamins DLK, Hong H, Cadieu CF, Solomon EA, Seibert D, DiCarlo JJ (2014) Performance-optimized hierarchical models predict neural responses in higher visual cortex. Proc Natl Acad Sci U S A 111:8619–8624.

Zhu Z, Xie L, Yuille AL (2016) Object Recognition with and without Objects. arXiv 1611.06596.

**ACKNOWLEDGEMENTS**

This research was supported by Intermediate and Senior Fellowships from the DBT/Wellcome Trust India Alliance to SPA (Grant #: 500027/Z/09/Z and IA/S/17/1/503081). HK was supported by a DST-CSRI postdoctoral fellowship from the Department of Science and Technology (DST).

**AUTHOR CONTRIBUTIONS**

PRT, HK & SPA designed the study, interpreted the results and wrote the manuscript. PRT & HK implemented foveation, trained neural networks and collected data.

**Supplementary material for**

**Human peripheral blur is optimal for object recognition**

RT Pramod[1][*][#], Harish Katti[2][*][#], & SP Arun[2][#]

[1]Department of Electrical Communication Engineering & [2]Centre for Neuroscience
Indian Institute of Science, Bangalore 560012

**CONTENTS**

# SECTION S1: OPTIMALITY ACROSS VIEWING DISTANCES

An object in a scene can undergo different amount of blur depending on the distance from which it is viewed. As a consequence, the foveated object has to be at an optimal distance from the viewer such that all relevant features can be extracted in high resolution for maximum performance. However, we do not know the optimal viewing distance for the ImageNet dataset. To address this issue, we trained multiple VGG-16 models each with a different viewing distance and on images with human-like foveation corresponding to that distance. We then chose the viewing distance (= 120 cm) corresponding to the best model to test object recognition performance with steeper or shallower blur profiles.

To verify that our results are not circular and that human-like foveal blur is optimal at other viewing distances as well, we selected another viewing distance that simulated an observer viewing the ImageNet images on a monitor placed 90 cm away and trained two additional VGG-16 models with steeper and shallower blur profiles (spatial decay factor of 2 and 0.5 respectively). In addition to these, we also trained a VGG-16 model trained on images with human-like blur profile. We then evaluated the test accuracy on the held out set of 50,000 images across 1000 categories and found that the model trained with human like foveation indeed had the highest accuracy (top 5 accuracy: 70.5% for steeper, 72.0% for human, and 70.1% for shallower blur across 1000 categories, viewing distance = 90 cm). Thus, human-like peripheral blur is across different viewing distances.

# SECTION S2: GENERALIZATION ACROSS DEEP NETWORK ARCHITECTURES

The results in the main text were reported on training a VGG-16 architecture on the ImageNet database. To confirm that this result is not specific to the choice of network or database, we trained a custom end-to-end CNN to recognize the presence of people in scenes chosen from the Microsoft common objects in context (MSCOCO) dataset (Lin et al., 2014).

**METHODS**

*Database.* The MSCOCO dataset version we used, consisted of 82079 training images and 40,137 test images. Images in this dataset were pre-annotated with bounding boxes for up to 80 different categories including people. We selected 44,665 images from MSCOCO that had instances of people and an equal number of negative examples drawn from 79 remaining non-person categories. In this manner we created a new set of 89,330 images half of which had people in them. Likewise, we created a test set of 42,786 images in which half the number of scenes had people in them. In scenes containing multiple annotated instances of people, we randomly chose only one instance so that each unique image was present only once in either the training set or the test set.

*CNN architecture.* We used a feed-forward architecture that is popular for binary classification task such as classifying the presence of dogs and cats in natural images (Parkhi et al., 2012). The network is different from VGG-16 (Chatfield et al., 2014) and contained three convolutional layers with 3x3 stride and 32, 32 and 64 channels respectively. Every convolutional layer was accompanied by ReLU activation and max-pooling over a 2x2 neighbourhood. The convolutional layers were flattened to a dense layer with 500 nodes and then to a single decision node. A drop-out factor of 0.5 was applied between the 500-node penultimate dense layer and the single decision node and sigmoid non-linearity was used to re-scale decision residuals to the range [0,1]. We used the binary cross-entropy loss metric and stochastic gradient descent optimizer.

*CNN training.* For each network, we started with randomly initialized weights and trained the network for binary person present/absent classification till accuracy on the held-out test set plateaued out or till 1000 epochs. All networks were defined and trained using python scripts and the Keras framework with NVIDIA TITAN-X/1080i GPUs. All the trained models were tested for generalization capabilities on a corresponding test set of 42,786 images in which half the number of scenes had people in them. (MSCOCO validation set).

*Foveation.* To create images with different levels of foveal blur, we used a similar procedure as the one used for VGG-16 (Chatfield et al., 2014) architecture based CNNs that we trained using Imagenet (Russakovsky et al., 2015). We first trained separate networks on images foveated with different viewing distances 30, 60, 90, 150 and 180 cm and found a general advantage of foveation over training with intact scenes. We then used novel spatial frequency decay factors that were either shallower or steeper than the human contrast sensitivity function, using a procedure similar to that employed for the VGG-16 based networks. We did this once for the 60cm viewing distance and once for the 180 cm viewing distance and compared the person classification accuracy with human like foveation at these viewing distances.(Lin et al., 2014)

**RESULTS**

Across all trained models, the network trained on images foveated according to the human peripheral blur function gave better performance on the test set containing 44,655 images than those trained with intact scenes (Table S1). Specifically, we obtained increases in accuracy of 1.2% and 2.2% with models trained on human-like foveated images over those trained on full resolution images for viewing distances of 60 cm and 180 cm respectively.

Thus, images foveated according to the human peripheral blur function yielded optimal recognition performance compared to other blur profiles.

| spatial decay factor | Test accuracy | |
| --- | --- | --- |
| | Viewing distance 60 cm | Viewing distance 180 cm |
| 4 | 60.8% | 62.3% |
| 2 | 60.9% | 62.0% |
| **1 (human)** | **63.2%** | **64.0%** |
| 0.5 | 63.2% | 62.5% |
| 0.25 | 62.5% | 62.9% |
| No foveation | 62.0% | 61.8% |

**Table S1. Person classification performance for varying peripheral blur.** Test accuracies are reported for viewing distances of 60 cm and 180 cm. Network trained on foveation level matching the human contrast sensitivity function (foveation spatial decay factor = 1) is highlighted in *red*. All models were trained to saturation and the best model for each spatial decay factor was chosen from the last 10 training epochs. Accuracies reported are calculated on the test set of 42,786 images that were not used for model training.

# SECTION S3: CAR DETECTION TASK

We performed an additional experiment to measure human performance on an inanimate category task. In this experiment, subjects detected the presence of cars. An example scene is shown in Figure S1A.

**METHODS**
All methods were identical to Experiments 1 & 2 except those detailed below.
*Images.* Like in the person detection experiment, we chose 120 images that contained cars and 120 images that had other objects (for example: dog, bird, boat, dustbin, post-box, bench, window, chair). The scenes without cars were the same as the no-person scenes in Experiment 2. All other aspects of the design are similar to the person detection task (Experiment 2).

**RESULTS**
Subjects were highly accurate in detecting the target object (average accuracy: 88.1%). Just like the animal and person tasks, categorization accuracy was similar across all levels of foveation except for steeply foveated images (median accuracy: 89.1% for full-resolution images, 88.9% for human foveated images, 89.9% for shallow and 88.1% for steeply foveated images; Figure S1B). We found similar trends in the reaction times, although these trends were not significant (average reaction times: 507 ms for full resolution images, 517 ms for shallow foveated images, 502 ms for human foveated images, and 528 ms for steeply foveated images; Figure S1C).

Thus, categorization performance in the car task was unaffected by blur.

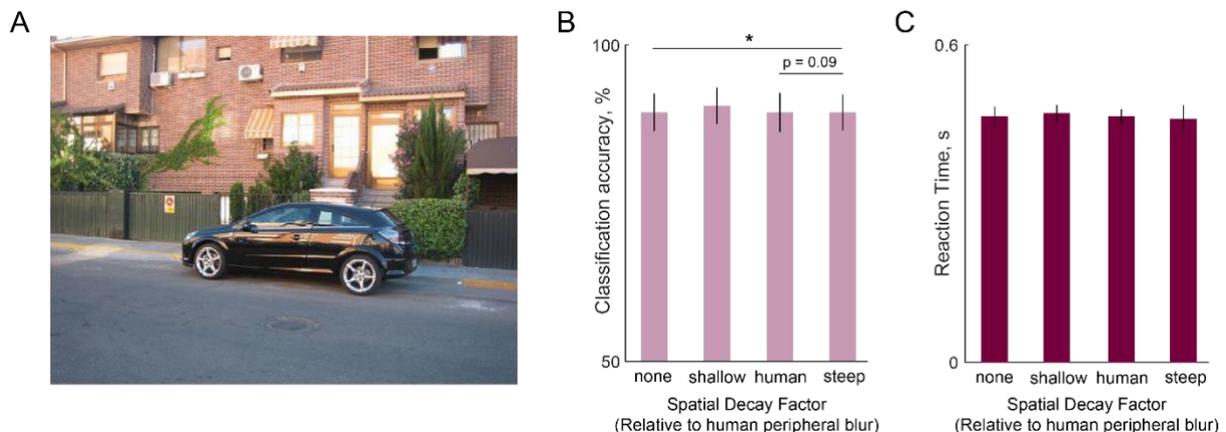

**Figure S1. Human behaviour on car detection.**
  (A) Example full resolution image used in the car detection task.
  (B) Bar plot of accuracy for different levels of peripheral blur. Error bars indicate s.e.m. calculated across images.
  (C) Same as in (B) but for reaction time on correct trials.

# SUPPLEMENTARY REFERENCES

Chatfield K, Simonyan K, Vedaldi A, Zisserman A (2014) Return of the Devil in the Details: Delving Deep into Convolutional Nets. In: Proceedings of the British Machine Vision Conference 2014, pp 6.1-6.12. British Machine Vision Association.

Lin T-Y, Maire M, Belongie S, Bourdev L, Girshick R, Hays J, Perona P, Ramanan D, Zitnick CL, Dollár P (2014) Microsoft COCO: Common Objects in Context. arXiv.

Russakovsky O, Deng J, Su H, Krause J, Satheesh S, Ma S, Huang Z, Karpathy A, Khosla A, Bernstein M, Berg AC, Fei-Fei L (2015) ImageNet Large Scale Visual Recognition Challenge. Int J Comput Vis 115:211–252.